\documentclass[12pt]{article}
\usepackage{graphicx}
\usepackage{amssymb}
\usepackage{amsmath}

\newcommand{\bear}{\begin{array}}  
\newcommand {\eear}{\end{array}}
\newcommand{\bea}{\begin{eqnarray}}   
\newcommand{\eea}{\end{eqnarray}}
\newcommand{\beq}{\begin{equation}}   
\newcommand{\eeq}{\end{equation}}
\newcommand{\bef}{\begin{figure}}  \newcommand 
{\eef}{\end{figure}}
\newcommand{\bec}{\begin{center}}  \newcommand 
{\eec}{\end{center}}

\begin{document}

\begin{titlepage}

\begin{flushright}
ICRR-Report-550\\
IPMU 09-0087 \\
\end{flushright}

\vskip 1.35cm


\begin{center}

{\large \bf
Effects of Dark Matter Annihilation on the \\Cosmic Microwave Background
}

\vskip 1.2cm

Toru Kanzaki$^a$,
Masahiro Kawasaki$^{a,b}$ and
Kazunori Nakayama$^a$

\vskip 0.4cm

{ \it $^a$Institute for Cosmic Ray Research,
University of Tokyo, Kashiwa 277-8582, Japan}\\
{\it $^b$Institute for the Physics and Mathematics of the Universe,
University of Tokyo, Kashiwa 277-8568, Japan}

\date{\today}

\begin{abstract} 
We study the effects of dark matter annihilation during and after the cosmic recombination epoch
on the cosmic microwave background anisotropy, taking into account the
detailed energy deposition of the annihilation products. 
It is found that a fairly stringent constraint on the annihilation cross section is imposed
for TeV scale dark matter masses from WMAP observations.
\end{abstract}



\end{center}
\end{titlepage}

\section{Introduction}

Although the existence of dark matter in the Universe is promising, 
its observational evidences rely only on its gravitational properties.
Despite tremendous efforts, particle physics natures of the dark matter remains unrevealed.

Many possible ways to detect the dark matter were proposed so far~\cite{Jungman:1995df}.
Direct detection experiments search for the dark matter scattering signatures off nucleons.
Indirect detection experiments search for characteristic cosmic-ray signals 
originated from dark matter annihilation inside and/or outside the Galaxy.
Recently the PAMELA satellite reported an excess of cosmic-ray positron flux 
around the energy $10$-100~GeV~\cite{Adriani:2008zr}, 
and it can be interpreted as a signal of dark matter annihilation.
Results of the Fermi satellite~\cite{Abdo:2009zk} on the cosmic-ray electron plus positron flux
seems to be consistent with the dark matter interpretation of the PAMELA excess.
If this scenario is correct, however, a fairly large annihilation cross section,
typically $\langle \sigma v \rangle \sim 10^{-24}$-$10^{-23}~{\rm cm^3s^{-1}}$ is required,
which is two- or three-orders of magnitude larger than the canonical value
($\langle \sigma v \rangle \simeq 3\times 10^{-26}~{\rm cm^3s^{-1}}$) reproducing the
correct dark matter abundance under the standard thermal freeze out scenario.
It was pointed out that the existence of dark matter with such a large annihilation cross section 
is constrained from many observations :
gamma-rays from Galactic center~\cite{Bertone:2008xr} or 
from extragalactic background~\cite{Kawasaki:2009nr,Profumo:2009uf},
neutrinos~\cite{Hisano:2008ah}, anti-protons~\cite{Cirelli:2008pk}, 
synchrotron radiation, 
and even from big-bang nucleosynthesis (BBN)~\cite{Jedamzik:2004ip}.
More or less, all of these constraints so far seem to be comparable to one another.
Some models can be killed by some observations, but there is a possible way out.
For example, the Galactic gamma-ray constraint is often very stringent but sensitive to the
density profile of the dark matter halo.
It can be evaded by assuming the flat cored profile.

In this paper we propose a more rigorous way to constrain dark matter annihilation cross section.
Although dark matter annihilates with each other throughout whole history of the Universe,
let us now focus on its effects around the recombination epoch.
Dark matter annihilation injects high energy particles into thermal plasma consisting of
photons, electrons, protons, neutral hydrogens and heliums.
High energy particles interact with background plasma and some fraction of their energy is used for
the ionization of neutral hydrogens.
Therefore, dark matter annihilation affects the recombination history of the hydrogen
resulting in higher residual ionization fraction ($x_e$) than the standard value.
Even a slight shift of $x_e$ can cause a drastic effect on the cosmic microwave background (CMB)
anisotropy, and hence the annihilation rate is constrained from the
precise measurements of the CMB anisotropy.

Similar analyses were carried out in several 
papers~\cite{Padmanabhan:2005es,Zhang:2006fr,Mapelli:2006ej,Natarajan:2008pk},
especially in connection with recent anomalous cosmic-ray positron/electron fluxes
\cite{Belikov:2009qx,Galli:2009zc,Huetsi:2009ex,Cirelli:2009bb,Slatyer:2009yq}.
Many of those analyses are based on the study in Ref.~\cite{Chen:2003gz} where 
effects of high-energy electron/photon energy injection due to late decaying particles on CMB 
were investigated using the extrapolation of the result of Ref.~\cite{Shull}
for the energy fraction going into the ionization, heating and excitation of the intergalactic medium,
which was valid only for low energy electron/photon injection.
Ref.~\cite{Kanzaki:2008qb} derived a formalism to extend this
to high-energy electron/photon energy injection, but there the Hubble expansion effect was neglected.
Therefore, we further extend the formalism of Ref.~\cite{Kanzaki:2008qb} to correctly handle the
Hubble expansion, and apply it to the study of the CMB constraint on the dark matter annihilation.
(See Ref.~\cite{Slatyer:2009yq} for a different approach to compute the 
details of the energy deposition.)

This paper is organized as follows.
In Sec.~\ref{sec:chi} we generalize the formulation of Ref.~\cite{Kanzaki:2008qb}
to include the effect of Hubble expansion
for calculating the energy deposition from high energy photons and electrons.
In Sec.~\ref{sec:CMB} we apply the results of Sec.~\ref{sec:chi}
to derive the effects of  dark matter annihilation on the CMB anisotropy,
and constrain the annihilation cross section.
In Sec.~\ref{sec:clump} we will mention the effect of dark matter clustering.
Sec.~\ref{sec:conc} is devoted to the conclusions.

\section{Energy deposition from dark matter annihilation :
computational methods} \label{sec:chi}

In this section we describe our method to calculate which fraction of the injected 
electron and photon energy goes into the ionization, heating and excitation of the intergalactic medium.
(The case of another injected particles, such as a muon, will be discussed in the next section.)
Our method is closely based on Ref.~\cite{Kanzaki:2008qb} where
the effect of Hubble expansion was neglected.
We extend the formalism to include the Hubble expansion.

Let $E_1,E_2,\dots, E_N$ be a discrete set of energies of particles
($E_i < E_j$ for $i<j$).
Suppose that an electron in the energy bin $i$ is injected at the redshift $z$.
This electron interacts with background particles and redistribute its energy toward
lower bins in the form of both electrons and photons.
Then similar energy degradation processes occur for each low energy electron/photon,
and again similar processes are repeated until $z=0$.
During this cascade processes, some of the initial injected energy is used for
the heating, excitation and ionization of the intergalactic medium.
In order to deal with this entire process, let us also introduce a discrete set of the
redshift, $z_1,z_2,\dots, z_m$ ($z_i<z_j$ for $i<j$).

We define $\chi^a_i(E_i,z_m,z_n),\chi^a_h(E_i,z_m,z_n)$ and $\chi^a_e(E_i,z_m,z_n)$ as
fractions of the initial energy $E_i$ at $z_m$ for an electron ($a=e$) or photon $(a=\gamma)$
which goes into ionization, heating and excitation at the redshift bin of $z_n$, respectively.
We also introduce $\chi^a_z(E_i,z_m,z_n)$ as a fraction of the initial energy lost by the Hubble expansion at $z_n$.
The remaining energy, which is not yet used for the above processes is denoted by
$\chi^a_r(E_i,z_m,z_n)$.
These $\chi^a_\alpha(E_i,z_m,z_n)~(\alpha=i,h,e,z,r)$ are the quantities which we want to know.
In order to evaluate them, the following probabilities are necessary.
$P^a(E_i,E_j,z_m)$ denotes the probability for an electron ($a=e$) or photon ($a=\gamma$)
with energy $E_i$ is transferred to another energy bin $E_j$ by scatterings at $z_m$.
The background electron/photons are scattered up 
through the same process, $a+b({\rm BG})$~($\{a,b\} = \{e,\gamma\}$)
and the possibility for this is represented as $P^{ab}_{\rm BG}(E_i,E_j,z_m)$.
The possibility for an electron/photon in energy bin $i$ redshifts away is denoted by
$P^a_H(E_i,z_m,z_{m-1})$.
There are two relations which can be easily understood,
\begin{gather}
	\sum_{i>j}P^a(E_i,E_j,z_m)+P_H(E_i,E_j,z_m)=1, \\
	P^a(E_i,E_j,z_m) = \sum_{b=e,\gamma}P^{ab}_{\rm BG}(E_i,E_i-E_j,z_m).
\end{gather}
Let us first consider the case $z_m=z_n$. In this case $\chi_\alpha(E_i,z_m,z_m)$ are given by
\begin{equation}
\begin{split}
	&\chi^a_{i,e,h}(E_i,z_m,z_m)=\sum_{j<i}\left[  P^a(E_i,E_j,z_m) \chi^a_{i,e,h}(E_j,z_m,z_m)
	\right.\\
	&~~~~~\left.+P^{ab}_{\rm BG}(E_i,E_j,z_m) \chi^b_{i,e,h}(E_j,z_m,z_m)\right] \frac{E_j}{E_i},
\end{split}
\end{equation}
\begin{equation}
\begin{split}
	&\chi^a_{z}(E_i,z_m,z_m)=\sum_{j<i}\left[  P^a(E_i,E_j,z_m) \chi^a_{z}(E_j,z_m,z_m)
	\right.\\
	&~~~~~\left.+P^{ab}_{\rm BG}(E_i,E_j,z_m) \chi^b_{z}(E_j,z_m,z_m)\right] \frac{E_j}{E_i} 
	+P^a_H(E_i,z_m,z_{m-1})\frac{\Delta_z E_i}{E_i},
\end{split}
\end{equation}
where $\Delta_z E_i= E_i[1-(1+z_{m-1})/(1+z_m)]$, and
\begin{equation}
\begin{split}
	&\chi^a_{r}(E_i,z_m,z_m)=\sum_{j<i}\left[  P^a(E_i,E_j,z_m) \chi^a_{r}(E_j,z_m,z_m)
	\right.\\
	&~~~~~\left.+P^{ab}_{\rm BG}(E_i,E_j,z_m) \chi^b_{r}(E_j,z_m,z_m)\right] \frac{E_j}{E_i} 
	+P^a_H(E_i,z_m,z_{m-1})\frac{E_i-\Delta_z E_i}{E_i}.
\end{split}
\end{equation}
The energy conservation is checked by $\sum_\alpha \chi_\alpha(E_i,z_m,z_m)=1$.
When $z_n < z_m$,  $\chi_\alpha(E_i,z_m,z_n)$ are evaluated by
\begin{equation}
\begin{split}
	&\chi^a_{\alpha}(E_i,z_m,z_n)=\sum_{j<i}\left[  P^a(E_i,E_j,z_m) 
	\chi^a_{\alpha}(E_j,z_m,z_n)
	\right.\\
	&~~~~~\left.+P^{ab}_{\rm BG}(E_i,E_j,z_m) 
	\chi^b_{\alpha}(E_j,z_m,z_n)\right] \frac{E_j}{E_i} \\
	&~~~~~+P^a_H(E_i,z_m,z_{m-1})
	\chi^a_{\alpha}\left(E_i-\Delta_z E_i,z_{m-1},z_n \right)
	\frac{E_i-\Delta_z E_i}{E_i}.
\end{split}
\end{equation}
The energy conservation is also satisfied 
($\sum_\alpha \chi_\alpha(E_i,z_m,z_n)=\chi_r(E_i,z_m,z_{n+1})$).
Therefore, once we know all $\chi_\alpha(E_j,z_l,z_k)$ with $E_j<E_i$ and $z_l\leq z_m$,
we can compute $\chi_\alpha(E_i,z_m,z_n)$ using the above equations.

The remaining task is to determine the transition probabilities :
$P^a(E_i,E_j,z_m)$, $P_H^a(E_i,z_m,z_{m-1})$ and $P_{\rm BG}^{ab}(E_i,E_j,z_m)$.
They are related to the collision frequency of elementary processes.
We take into account all the relevant interactions.
For an electron,
the excitation and ionization of and collision with the hydrogen atom, 
Coulomb collision with the background electrons, 
and inverse Compton scattering with CMB photons are considered.
For a photon,
photoionization of the hydrogen atom, Compton scattering with background electrons,
pair cration in nuclei,
photon-photon scattering and double photo pair creation with CMB photons are considered.
For simplicity, we neglect $^4$He and assume that all baryons are H.

Since the energy loss processes contain both ``continuous loss'' where
the energy loss per collision is smaller than the energy bin $\Delta E_i$, and ``discrete loss''
where the energy loss is significant, we need to distinguish them for a numerical purpose.
The collision frequency for a discrete loss can be given by
\begin{equation}
	\nu_d^{ab}(E_i,E_j)=n_tv_p \frac{d\sigma^{ab}(E_i)}{dE_j}\Delta E_j,
\end{equation}
where $n_t$ is the number density of the target particle, $v_p$ is the particle velocity, and
$d\sigma^{ab}(E_i)/dE_j$ is the differential cross section.
The collision frequency for a continuous loss is expressed as
\begin{equation}
	\nu_c^a (E_i,E_{i-1})=\frac{1}{\Delta E_i}\left[ \frac{-dE}{dt} \right] _c.
\end{equation}
Similarly, we define the ``collision'' frequency for the Hubble expansion as
\begin{equation}
	\nu_H^a (E_i)=\frac{1}{\Delta_z E_i}\left[ \frac{-dE}{dt} \right] _H =
	\frac{H(z_m)E_i}{\Delta_z E_i},
\end{equation}
where $H(z)$ is the Hubble parameter at the redshift $z$.
Using these quantities, transitions probabilities are expressed as
\begin{eqnarray}
	P^a(E_i,E_{i-1},z_{m})=\frac{\nu_c^a (E_i,E_{i-1})+\nu_d^a (E_i,E_{i-1})}
	{\nu_c^a (E_i,E_{i-1})+\sum_{k}\nu_d^a (E_i,E_{k}) +\nu_H^a (E_i)} \nonumber , \\
	P^a(E_i,E_{j},z_{m})=\frac{\nu_d^a (E_i,E_{j})}
	{\nu_c^a (E_i,E_{i-1})+\sum_{k}\nu_d^a (E_i,E_{k}) +\nu_H^a (E_i)} \nonumber , \\
	P_H^a(E_i,z_{m},z_{m-1})=\frac{\nu_H^a (E_i)}
	{\nu_c^a (E_i,E_{i-1})+\sum_{k}\nu_d^a (E_i,E_{k}) +\nu_H^a (E_i)} \nonumber ,
\end{eqnarray}
where $\nu^a_d = \nu_d^{ae}+\nu_d^{a\gamma}$.
We also obtain 
\begin{equation}
\begin{split}
	P_{\rm BG}^{ab}(E_i,E_i-E_j,z_{m})
	=\frac{\nu_d^{ab} (E_i,E_j)}
	{\nu_c^a (E_i,E_{i-1})+\sum_{k}\nu_d^a (E_i,E_{k}) +\nu_H^a (E_i)}.
\end{split}
\end{equation}
The elementary processes contributing to $P_{\rm BG}^{ab}$ are as follows :
collisional ionization of the hydrogen atom for ($a=e,b=e$),
inverse Compton scattering for ($a=e,b=\gamma$),
photoionization, Compton scattering, photon-matter pair creation, double photon pair creation
for ($a=\gamma, b=e$), and
photon-photon scattering for ($a=\gamma, b=\gamma$).
All the relevant cross sections and energy loss rates are summarized in Ref.~\cite{Kanzaki:2008qb},
and we do not repeat here.


\begin{figure}[t]
 \begin{center}
   \includegraphics[width=0.7\linewidth]{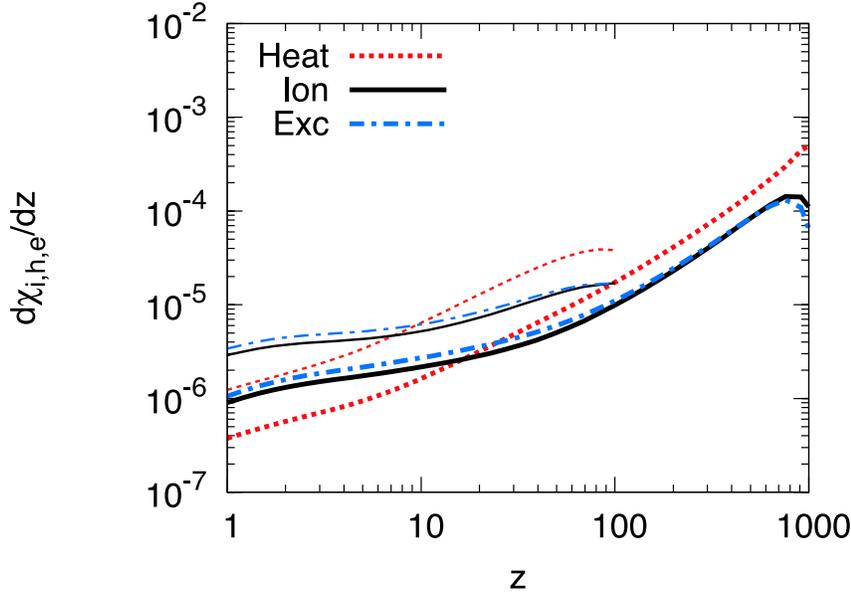} 
   \caption{ The differential ionization, heating and excitation fractions 
	   $d\chi^{(e)}_{i,h,e}(E,z',z)/dz$ are shown by red-dotted, black-solid and dot-dashed-blue lines,
	   respectively. Thick (thin) lines correspond to $z'=1000~(100)$, and we have taken
	   $E=1$~TeV. }
   \label{fig:dchidz}
 \end{center}
\end{figure}


The above methods allow us to compute $\chi^a_\alpha(E_i,z_m,z_n)$.
In Fig.~\ref{fig:dchidz} the differential ionization, heating and excitation fractions 
$d\chi^{(e)}_{i,h,e}(E,z',z)/dz$ are shown by red-dotted, black-solid and dot-dashed-blue lines,
respectively. Thick (thin) line corresponds to the injection redshift
$z'=1000~(100)$, and we have taken $E=1$~TeV.
Here no reionization from astrophysical sources is assumed for simplicity.\footnote{
	In this calculation, the background ionization history is fixed and 
         the extra contribution to the ionization is regarded as a small perturbation.
	We have checked that this approximation does not alter the following result at all,
	since otherwise the resultant cosmology significantly deviates from the standard one 
	and already excluded.
}
Now we are in a position to
accurately calculate the effects of dark matter annihilation on the CMB anisotropy,
as we will see in the next section.


\begin{figure}[t]
 \begin{center}
   \includegraphics[width=0.7\linewidth]{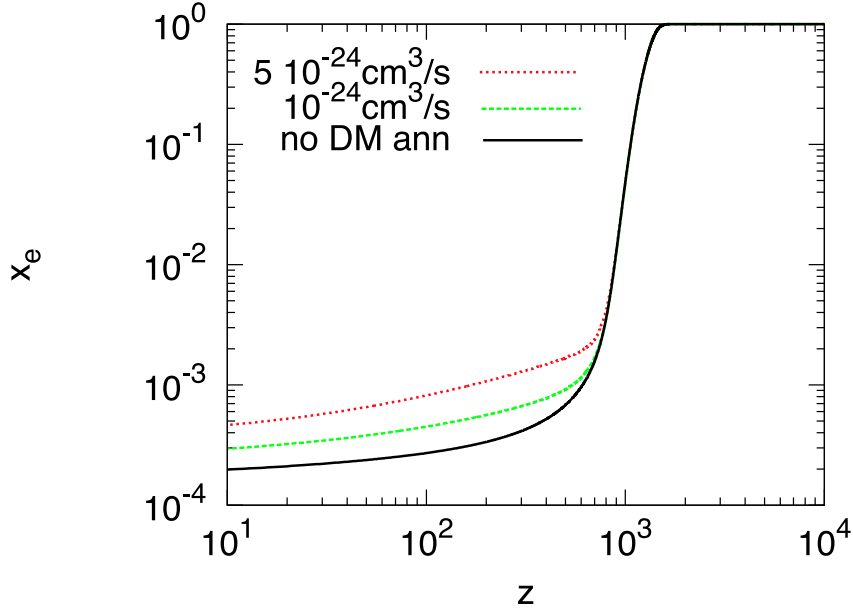} 
   \caption{ History of the ionization fraction as a function of redshift. 
   	The black solid line corresponds to the standard recombination history, without 
	dark matter annihilation effects. Also shown are the cases of dark matter with annihilation cross
	section $\langle \sigma v\rangle_{e^+e^-}=10^{-24}$ and $5\times 10^{-24}~
	{\rm cm^3 s^{-1}}$ with the dark matter mass $m_\chi$=1~TeV.}
   \label{fig:recomb}
 \end{center}
\end{figure}


\section{Effects of dark matter annihilation on the CMB anisotropy} \label{sec:CMB}

Now let us apply the results of previous section to the calculation of CMB anisotropy. 
In order to take into account the effects of extra ionization and heating source from dark matter
annihilation, we should add the following terms for the evolution of the ionization fraction
($x_e$) and the gas temperature $(T_b)$, 
\begin{equation}
\begin{split}
	-\left[ \frac{dx_e}{dz}\right]_{\rm DM}= &\sum_F \int_z \frac{dz'}{H(z')(1+z')}
	\frac{n_{\chi}^2(z') \langle \sigma v\rangle_F}{n_H(z')} 
	 \frac{m_\chi}{E_{\rm Ry}} \frac{d\chi_{i}^{(F)} (E,z',z)}{dz}, \label{dxedz}
\end{split}
\end{equation}
where $E_{\rm Ry}=13.6$~eV is the Rydberg energy,
$m_\chi$ and $n_\chi$ are the mass and number density of the dark matter particle,
$n_H$ is the number density of the hydrogen atom.
\begin{equation}
\begin{split}
	-\left[ \frac{dT_b}{dz}\right]_{\rm DM}= &\sum_F \int_z \frac{dz'}{H(z')(1+z')}
	\frac{2n_{\chi}^2(z') \langle \sigma v\rangle_F}{3n_H(z')}
	m_\chi \frac{d\chi_{h}^{(F)} (E,z',z)}{dz}. \label{dTbdz}
\end{split}
\end{equation}
Here we have defined
\begin{equation}
\begin{split}
	\frac{d\chi_{i,h}^{(F)} (E,z',z)}{dz}= \int dE~\frac{E}{m_\chi}
	\left[ \frac{dN_F^{(e)}}{dE}\frac{d\chi_{i,h}^{(e)}(E,z',z)}{dz}
		+ \frac{dN_F^{(\gamma)}}{dE}\frac{d\chi_{i,h}^{(\gamma)} (E,z',z)}{dz}\right],
	\label{dchiF}
\end{split}
\end{equation}
where $dN^{(e,\gamma)}_F/dE$ denotes the spectrum of the electron and photon
produced per dark matter annihilation into the mode $F$, and
$\langle \sigma v\rangle_F$ denotes the annihilation cross section into that mode.
We have included these terms in the RECFAST code~\cite{Seager:1999bc},
which is implemented in the CAMB code~\cite{Lewis:1999bs} for calculating the CMB anisotropy.
Here and hereafter, we fix the cosmological parameters to the WMAP five year best fit values
\cite{Dunkley:2008ie}.
The reionization optical depth is also fixed to be the best fit value and need not be reevaluated 
when the dark matter annihilation effect is included, 
since it depends only on the reionization history at low-redshift.
It is noted that the energy integral in (\ref{dchiF}) for given final states $F$
can be performed before solving the evolution equation
once we have tables of $d\chi^{(e)}_{i,h}/dz$ and $d\chi^{(\gamma)}_{i,h}/dz$,
as long as the produced particle $F$ decays so quickly as not to lose energy 
by the interaction with surrounding medium, which is a valid assumption
for any unstable standard model particle.

Fig.~\ref{fig:recomb} shows the evolution of the ionized fraction with and without 
the effects of dark matter annihilation.
The black solid line corresponds to the standard recombination history, without 
dark matter annihilation effects. Also shown are the cases of dark matter with annihilation cross
section $\langle \sigma v\rangle_{e^+e^-}=10^{-24}$ and 
$5\times 10^{-24}~{\rm cm^3 s^{-1}}$ with $m_\chi$=1~TeV.
It is clearly seen that energy injection from dark matter annihilation works as an 
extra source of ionization.
Then it is not hard to imagine that this modified recombination history affects the spectrum of
CMB anisotropy, since large $x_e$ implies large optical depth.

\begin{figure}
 \begin{center}
   \includegraphics[width=0.5\linewidth]{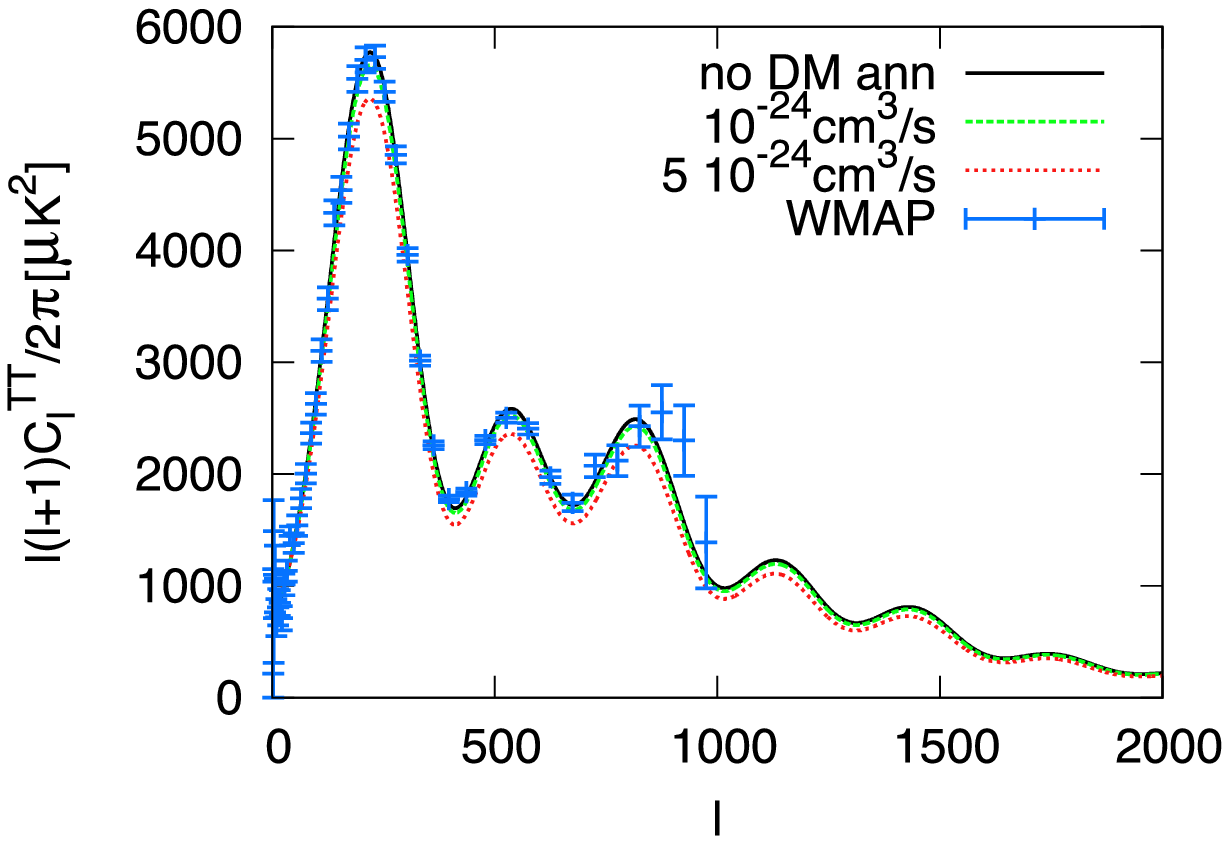} 
   \includegraphics[width=0.5\linewidth]{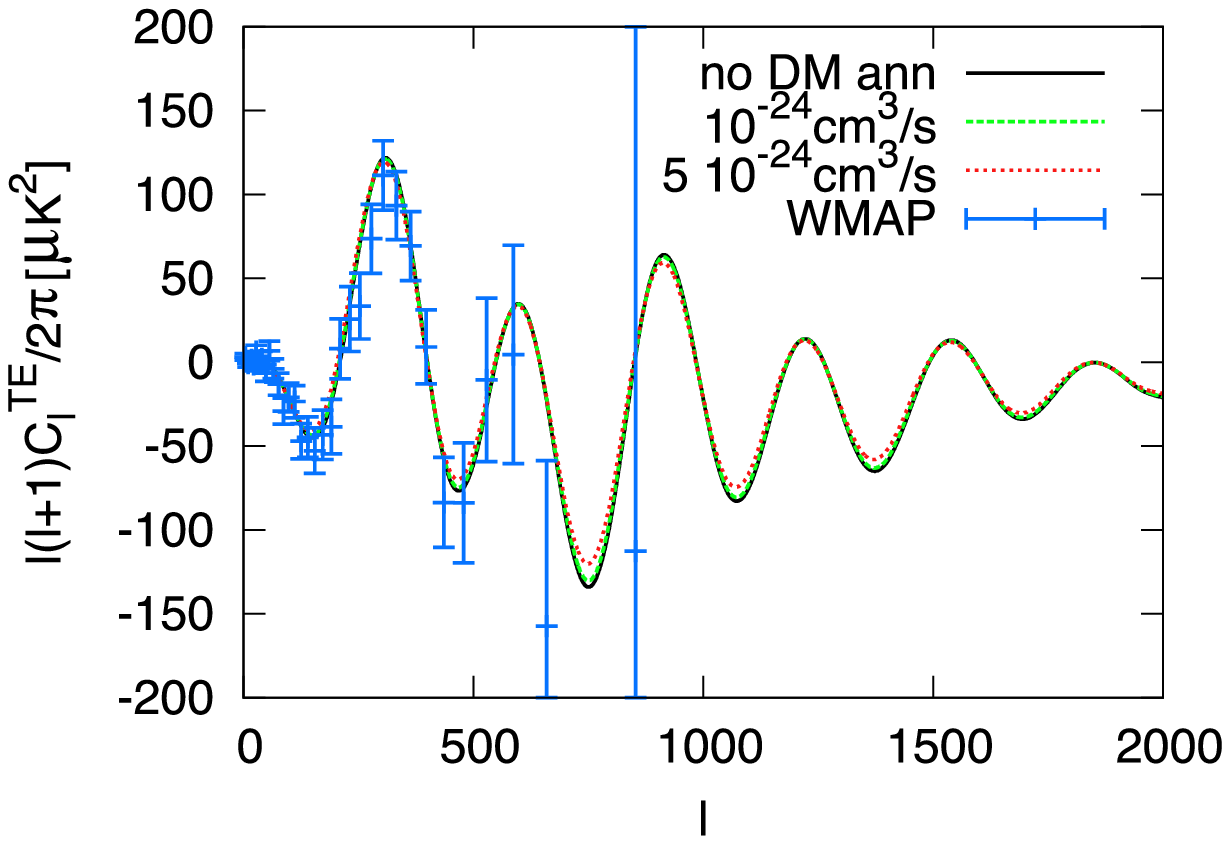} 
   \includegraphics[width=0.5\linewidth]{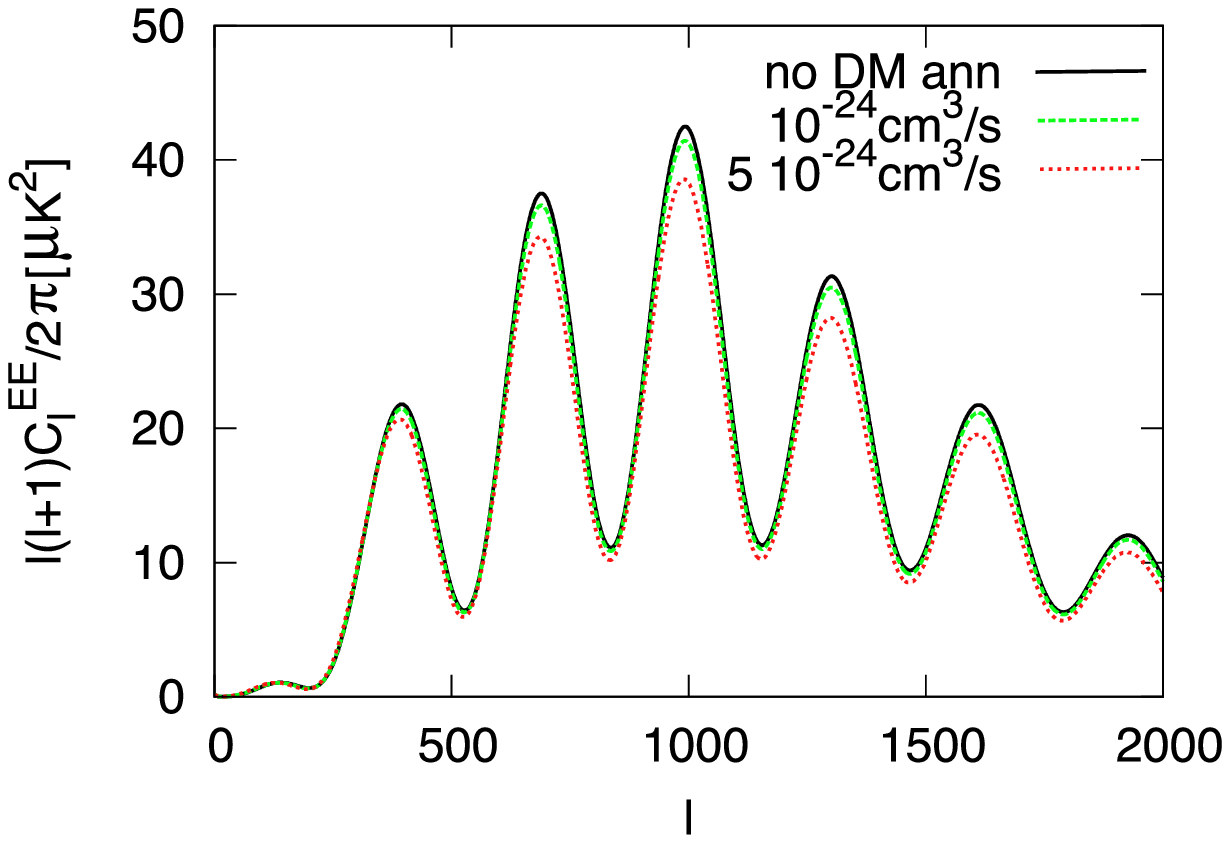} 
   \caption{ (Top) $TT$ spectrum of the CMB anisotropy. 
   	The black solid line shows the WMAP5 best fit curve without dark matter annihilation.
	Also shown are the cases of dark matter with annihilation cross
	section $\langle \sigma v\rangle_{e^+e^-}=10^{-24}$ and $5\times 10^{-24}~
	{\rm cm^3 s^{-1}}$ with mass of $m_\chi=$1~TeV. 
	(Middle) The same, but for $TE$ spectrum. (Bottom) The same, but for $EE$ spectrum.  }
   \label{fig:ClTT}
 \end{center}
\end{figure}

Fig.~\ref{fig:ClTT} show the multipole coefficient of the CMB anisotropy ($C_\ell$)
of $TT,TE$ and $EE$ modes, respectively.
The cases of $\langle \sigma v\rangle_{e^+e^-}=10^{-24}$ and 
$5\times 10^{-24}~{\rm cm^3 s^{-1}}$ with $m_\chi$=1~TeV are also shown.
It is seen that as the dark matter annihilation cross section is increased,
the spectrum is more suppressed and it will deviates from the observed data.


\begin{figure}[t]
 \begin{center}
   \includegraphics[width=0.6\linewidth]{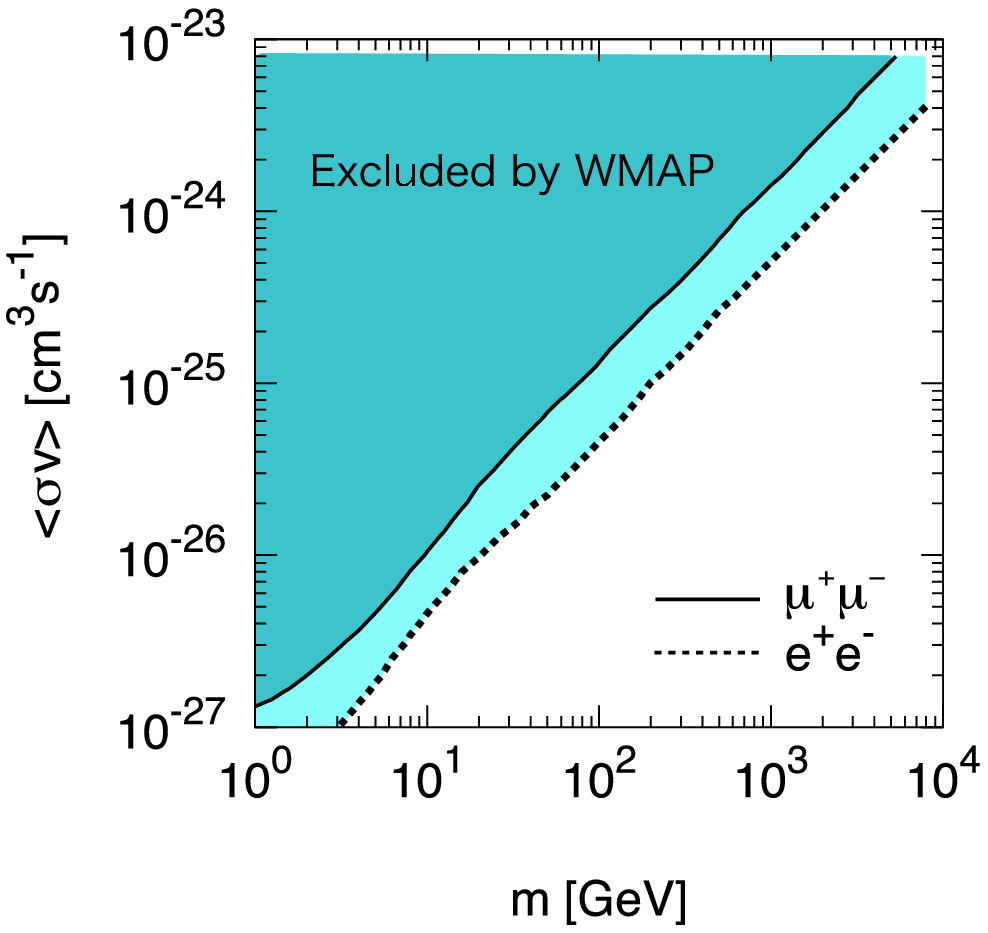} 
   \caption{ Constraint on the dark matter annihilation cross section with 99\% C.L.
   	into $e^+e^-$ (dashed) and $\mu^+ \mu^-$ (solid)
	as a function of the dark matter mass.
   }
   \label{fig:chi}
 \end{center}
\end{figure}

In Refs.~\cite{Huetsi:2009ex,Cirelli:2009bb}, the optical depth was calculated
for comparison with the CMB observation.
In this paper, we have performed a $\chi^2$ analysis in order to compare the results quantitatively
with the real data, using the WMAP likelihood code~\cite{LAMBDA}.
In Fig.~\ref{fig:chi}, resulting 
99\% C.L. constraints on the dark matter annihilation cross section are shown.
The solid (dashed) line corresponds to the case where dark mater annihilates into
$\mu^+ \mu^-$ ($e^+e^-$).
It is seen that CMB measurements gives stringent upper bound on the annihilation rate.
Note that, however, in this analysis we have fixed cosmological parameters as WMAP five year
best fit values.
It may be expected that there are some degeneracies between dark matter annihilation cross section
and other cosmological parameters, as shown in Ref.~\cite{Padmanabhan:2005es,Galli:2009zc}.
Thus the bound will be somewhat loosen when the full Monte Carlo analysis is performed.
We will give more detailed analysis on this subject elsewhere.


\section{Effects of dark matter clustering} \label{sec:clump}

So far, we have assumed that the dark matter distributes homogeneously in the whole Universe.
Of course, this is not true especially for the low redshift Universe, since the structure formation
of the dark matter begins to be important.
The dark matter annihilation rate per volume is proportional to $n_\chi^2$, 
and hence the clustering of the dark matter roughly replaces this with the average of the 
density squared, $\langle n_\chi^2 \rangle$, which can be much larger than the
homogeneous part, $\bar n_\chi^2$.
The importance of this effect was recognized in Ref.~\cite{Ullio:2002pj} for the calculation of 
the diffuse gamma-rays from dark matter annihilation 
(see Refs.~\cite{Kawasaki:2009nr,Profumo:2009uf} for recent studies).
In the context of cosmic reionization, the effect of dark matter clustering was pointed out in
Refs.~\cite{Natarajan:2008pk,Belikov:2009qx,Huetsi:2009ex,Cirelli:2009bb}.

The effect of dark matter clustering is represented by the enhancement factor $\Delta^2(z)$,
given by~\cite{Ullio:2002pj} 
\begin{equation}
	\Delta^2(z) = \frac{1}{\bar \rho_\chi}\int dM M \frac{dn(z)}{dM}\Delta_M^2(M,z),
	\label{Delz}
\end{equation}
where
\begin{equation}
	\Delta_M^2(z) = \frac{\Delta_{\rm vir}(z)}{3} 
	\frac{I_2(x_m)x_m^3}{I_1^2(x_m)}.
\end{equation}
Here $dn(z)/dM$ is the comoving number density of collapsed objects with mass $M$,
which can be evaluated by the Press-Schechter theory~\cite{Press:1973iz,Sheth:1999su}.
$\Delta_{\rm vir}(z)$ denotes the virial overdensity which, in the $\Lambda$CDM Universe,
is given by~\cite{Bryan:1997dn}
\begin{equation}
	\Delta_{\rm vir}(z) = \frac{18\pi^2+82y-39y^2}{\Omega_m(z)}, 
\end{equation}
where $\Omega_m(z)=\Omega_m(1+z)^3/[\Omega_\Lambda +\Omega_m(1+z)^3]$
and $y=\Omega_m(z)-1$.
$I_n(x_m)$ is defined as $I_n(x_m)=\int^{x_m} dx x^2 d^n(x)$
with $d(x)$ denoting the density profile of a dark mater halo, as
$\rho(r)=\tilde \rho d(r/a)$ with typical halo size $a$.
Finally, $x_m = R_{\rm vir}/a$ with $R_{\rm vir}$ being the virial radius of the objects with mass $M$ :
\begin{equation}
	\frac{4\pi}{3}R_{\rm vir}^3\Delta_{\rm vir}(z) \bar \rho_m(z) = M.
\end{equation}
For a given halo with mass $M$, the typical halo radius $a$ may have some 
correlation with the virial radius $R_{\rm vir}(M)$.
We adopt a model of Ref.~\cite{Bullock:1999he}.

It is also noticed that the mass integral in Eq.~(\ref{Delz}) has a lower cut $(M_{\rm cut})$
below which the dark matter does not cluster, depending on the kinetic decoupling
temperature of the dark matter particle~\cite{Loeb:2005pm,Profumo:2006bv},
and this choice determines the redshift at which the clustering effects become important,
as explicitly shown in Ref.~\cite{Kawasaki:2009nr}.
We assume $M_{\rm cut}=10^{-10}M_\odot$ in the following.
Hereafter we adopt the Moore density profile~\cite{Moore:1999gc}, where
$d(x)=x^{-1.5}(1+x^{1.5})^{-1}$.
It is noticed that the clustering effect is more significant for halo density profile with steep core,
such as the Moore's one.
In order to include the enhancement effect, we only have to replace $n_\chi^2 (z')$
with $n_\chi^2(z') [1+\Delta^2(z')]$ in Eqs.~(\ref{dxedz}) and (\ref{dTbdz}).

\begin{figure}[t]
 \begin{center}
   \includegraphics[width=0.7\linewidth]{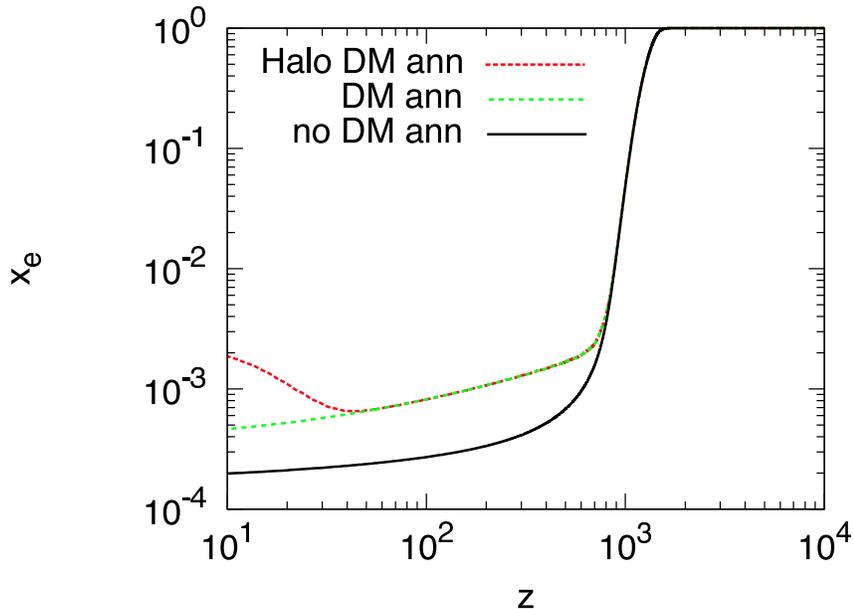} 
   \caption{ Evolution of the ionization fraction when no effects of dark matter annihilation
   	is taken into account (black solid), when dark matter with $\langle \sigma v \rangle
	=5\times 10^{-24}~{\rm cm^3s^{-1}}$ and $m_\chi=1$~TeV 
	are included (green dashed),
	and when the effects of dark matter clustering is also included (red dotted).
	We have assumed that the main annihilation mode is into $e^+e^-$.
   }
   \label{fig:zxehalo}
 \end{center}
\end{figure}

Fig.~\ref{fig:zxehalo} shows an evolution of the ionization fraction when no effects of dark matter annihilation is taken into account (black solid), 
when dark matter with $\langle \sigma v \rangle=5\times 10^{-24}~{\rm cm^3s^{-1}}$ 
and $m_\chi=1$~TeV are included (green dashed),
and when the effects of dark matter clustering is also included (red dotted).
In the last case, we have multiplied $\Delta^2(z)$ by a factor 10 in order to clearly see the
effects of dark matter clustering.
In this figure we have assumed that the main annihilation mode is into $e^+e^-$.
It is clearly seen that the dark matter clustering at low redshift enhances the annihilation rate
and increases the ionization fraction.
This effect, however, is found to be subdominant for the CMB anisotropy,
since the main effect on the CMB comes from the energy deposition around the recombination epoch
$(z\sim 1000)$.
A situation in which the ionization fraction is largely affected by the 
clustering effect at the low redshift is excluded even if the clustering effect is ignored.
Therefore, for the purpose of constraining dark matter annihilation models
from the observation of CMB anisotropy,
taking only the homogeneous dark matter component is sufficient,
in agreement with the statement in Ref.~\cite{Huetsi:2009ex}.


\section{Conclusions and discussion} \label{sec:conc}

We have studied the effects of dark matter annihilation on the CMB anisotropy,
motivated by the fact that recent measurements of anomalous comic-ray positron/electron fluxes
can be explained by the contribution from dark matter annihilation with
fairly large annihilation cross section.
In contrast to many preceding works, we have taken into account all the relevant 
energy loss processes of electrons and photons and estimated
how the injected energy goes into ionization, excitation and heating of atoms.

This method is applied to the calculation of CMB anisotropy, and it is found that
WMAP5 results give stringent bound on the annihilation cross section.
This is much stronger than that from BBN~\cite{Jedamzik:2004ip} and
other astrophysical bounds.
Gamma-ray observations also give stringent bound~\cite{Bertone:2008xr,Kawasaki:2009nr}, 
but it has rather large uncertainty due to the lack of the understandings of the dark matter halo profile.
Taking in mind all the astrophysical uncertainties on these cosmic-ray signatures, 
the CMB constraint given in the present paper is said to be robust :
the physics of CMB anisotropy is well understood from both theoretical and observational sides.
Moreover, the forthcoming Planck satellite will give more precise data of the CMB anisotropy
including polarization toward much smaller scales than the WMAP,
and hence the bound is expected to be significantly improved.
Furthermore, the heating of the intergalactic medium by the dark matter annihilation 
may have an impact on the 21cm observations~\cite{Furlanetto:2006wp,Natarajan:2009bm}.
We will pursue these issues in a separate publication.


\section*{Acknowledgment}

K.N. would like to thank T.~Sekiguchi for useful discussion.
He also would like to thank the Japan Society for the Promotion of Science for financial support.
This work is supported by Grant-in-Aid for Scientific research from the Ministry of Education,
Science, Sports, and Culture (MEXT), Japan, No.14102004 (M.K.)
and also by World Premier International
Research Center Initiative (WPI Initiative), MEXT, Japan.


{}

\end{document}